\documentclass[preprint,showpacs,nofootinbib,amsmath,amssymb]{revtex4}

\usepackage[dvips]{graphicx}
\usepackage{amsmath,amsthm,amssymb}

\begin{document}
\title{de Sitter Radiation and Backreaction in Quantum Cosmology}

\author{Je-An Gu}
\email{jagu@ntu.edu.tw}
\affiliation{Leung Center for Cosmology and Particle
Astrophysics, National Taiwan University, Taipei 10617, Taiwan}

\author{Sang Pyo Kim}
\email{sangkim@kunsan.ac.kr}
\affiliation{Department of Physics, Kunsan National University,
Kunsan 573-701, Korea}
\affiliation{Institute of Astrophysics, Center for Theoretical Physics, Department of Physics, National Taiwan University,
Taipei 106, Taiwan}

\author{Che-Min Shen}
\email{f97222024@ntu.edu.tw}
\affiliation{Department of Physics and Leung Center for
Cosmology and Particle Astrophysics, National Taiwan
University, Taipei 10617, Taiwan}

\begin{abstract}
We explore the quantum cosmology description of the de Sitter
(dS) radiation and its backreaction to dS space, inherent in
the wave function of the Wheeler-DeWitt equation for pure
gravity with a cosmological constant. We first investigate the
quantum Friedmann-Lemaitre-Robertson-Walker cosmological model
and then consider possible effects of inhomogeneities of the
universe on the dS radiation. In both the cases we obtain the
modified Friedmann equation, including the backreaction from
spacetime fluctuations, and the quantum-corrected dS
temperature. It is shown that the quantum correction increases
the dS temperature with the increment characterized by the
ratio of the dS scale to the Planck scale.
\end{abstract}


\maketitle


\textit{Introduction}.---
The de Sitter (dS) space is essential in cosmology and in the
study of the quantum aspects of gravity. In cosmology the
inflation, the early-time accelerating expansion that is nearly
dS, plays an important role in giving the initial big bang and the
favorable initial condition of our universe. In particular it
generates the seeds of the cosmic structures, meanwhile giving
a rather flat, homogeneous and isotropic universe that is
simple to study. In addition, the late-time accelerating
expansion may eventually approach dS, since the $\Lambda$CDM
model with a positive cosmological constant for driving the
cosmic acceleration fits the current observational results
well.

The dS radiation, as the Hawking radiation for black holes,
entails quantum theory since it is a consequence of the unitary
inequivalence of the in-vacuum and the out-vacuum for quantum
field excitations. Though the dS radiation can be understood
within the quantum field theory on a classical spacetime, it
may also be studied in a more fundamental framework, quantum
cosmology or quantum gravity. One may thus raise a question
about the physics involved in the quantum cosmological model
for the dS radiation. In quantum cosmology prescribed by the
Wheeler-DeWitt (WDW) equation from the Hamiltonian formulation
of general relativity \cite{DeWitt:1967yk}, the universe is
described by a wave function of the spacetime geometry and the
matter fields, and so is the dS radiation. Accordingly, the dS
radiation, as well as its backreaction to dS space, should be
inherent in the quantum cosmology description of the dS
universe.

The quantum correction to the gravitational field equations in
quantum cosmology has been investigated by
Banks\,\cite{Banks:1985} and further elaborated by Brout et al
\cite{Brout:1987ya,Brout:1987tq}. The WDW equation for gravity
coupled to matter fields has two mass scales: the Planckian
mass scale for the gravity part and the mass or energy scale
for the matter fields. Therefore, the WDW equation may have the
Born-Oppenheimer approximation, used for an atom of heavy
nucleus and electrons, such that the heavy gravitational part
separates from the light matter part in the approximate
solution for the WDW equation and the gravitational part leads
to the semiclassical Einstein equation as the Hamilton-Jacobi
equation \cite{Banks:1985,Brout:1987ya,Brout:1987tq}. However,
the Banks-Born-Oppenheimer (BBO) approximation cannot be
employed for pure gravity with a cosmological constant because
the gravity is the only scale in the WDW equation. In this
case, one may use the de Broglie-Bohm interpretation of the
wave function, in which the oscillatory wave packet is peaked
along a new trajectory governed both by the classical potential
and the quantum potential
\cite{Holland:1993ee,Holland:1993book}. The dS spacetime
emerges from the oscillatory wave packet. Further the motion of
matter in imaginary time in classically forbidden
configurations of gravity can be related to an inverse
temperature \cite{Brout:1987ya,Brout:1987tq}.

In this paper we investigate the dS radiation and its
backreaction in quantum cosmology with pure gravity,
i.e., where we consider solely the spacetime metric in the
field content and explore the intrinsic dS radiation as the
gravitational fluctuations. We will firstly investigate the
case with a homogeneous and isotropic, i.e.\ simply
time-dependent, metric. We will then consider the possible
effects from the spatial inhomogeneities of the metric.
In both the cases we obtain the modified Friedmann equation and
thereby the modified dS temperature with the quantum correction
from the wave function that encloses the backreaction of
the dS radiation. For comparison, in the following we start
with the framework of the quantum field theory on classical
spacetime before presenting our quantum cosmology analysis.

\textit{de Sitter radiation in classical spacetime}.---
Here we treat dS space as a classical background and disregard
the backreaction of the dS radiation. Since Gibbons and Hawking
proposed the dS radiation \cite{Gibbons:1977mu} soon after
discovering the black hole radiation \cite{Hawking:1974sw},
there have been several derivations of the dS temperature
presented \cite{Spradlin:2001pw,Birrell:1982ix,Parikh:2002qh}.
In the following we invoke the correspondence between the
inverse temperature and the Euclidean time period for the
derivation.
We consider the classical Friedmann equation in the Euclidean
time for a closed universe ($k=1$) in unit of $c = \hbar =1$,
\begin{equation}
- \left( \frac{a^{\prime}}{a} \right)^2 + \frac{k}{a^2}
= \frac{\Lambda}{3} \equiv H_{\Lambda}^2  \, ,
\end{equation}
where the prime is the derivative with respect to the Euclidean
time and $\Lambda$ is the cosmological constant. This equation
gives the inverse temperature as a period
\cite{Brout:1987ya,Brout:1987tq}
\begin{equation}
\beta_0 \equiv \frac{1}{k_B T_0}
= 2 \int_{-1/H_{\Lambda}}^{1/H_{\Lambda}} \frac{da}{\sqrt{1-H_{\Lambda}^2 a^2}}
= \frac{2\pi}{H_{\Lambda}} \, .
\end{equation}
As a result, the dS temperature is given by
\begin{equation} \label{eq:dS-Temperature0}
T_0 = \frac{H_{\Lambda}}{2\pi k_B } \, .
\end{equation}

\textit{de Sitter radiation in quantum cosmology with
homogeneous metric perturbations}.---
Now we consider a system of pure gravity and a
cosmological constant with the action
\begin{equation}
S = -\frac{1}{16\pi G} \int d^4 x \sqrt{-g}
\left( R - 2 \Lambda \right) .
\end{equation}
In the following we derive the modified Friedmann equation
\cite{Kim:1996ae,Kim:1997te} and thereby the modified dS
temperature from the WDW equation for a homogeneous
and isotropic universe described by the closed Robertson-Walker
metric,
\begin{equation}
ds^2 = - N^2(t)dt^2 + a^2(t)d\Omega_3^2 \, .
\end{equation}
Substituting this metric into the action gives
\begin{equation}
S = -\frac{3}{8\pi G} \int dt \, a^3 \left[ \frac{\dot{a}^2}{Na^2}
+ N \left( -\frac{k}{a^2} + \frac{\Lambda}{3} \right) \right] ,
\end{equation}
where the overdot denotes the time derivative. The
WDW equation from the super-Hamiltonian constraint for the
Friedmann-Lemaitre-Robertson-Walker (FLRW) cosmological model
reads
\begin{equation}
\left[ -\frac{1}{2}G\frac{\partial^2}{\partial a^2}
+ \frac{9}{32\pi^2G}a^4\left(\frac{k}{a^2}-\frac{\Lambda}{3}\right)
\right] \Psi (a) = 0 \, ,
\end{equation}
where $\Psi(a)$ is the wave function of the universe.

We follow the de Broglie-Bohm interpretation for an oscillatory
wave packet \cite{Holland:1993ee,Holland:1993book}, in which
the wave packet is peaked along a trajectory guided not only by
the classical potential but also by the quantum potential. For
that purpose, one writes
\begin{eqnarray}
\Psi(a) = F(a) \exp [ iS(a)],
\end{eqnarray}
and obtains the semiclassical Einstein
equation from the real part of the WDW equation,
i.e.\ the Hamilton-Jacobi equation:
\begin{equation}
\frac{1}{2}G\left(\frac{\partial S}{\partial a}\right)^2
+ \frac{9}{32\pi^2G}a^4\left(\frac{k}{a^2}-\frac{\Lambda}{3}\right)
+ V_q(a) = 0 \, , \label{HJ eq}
\end{equation}
where the quantum potential
\begin{equation}
V_q(a) = - \frac{G}{2F} \frac{\partial^2 F}{\partial a^2} \, .
\end{equation}
On the other hand, the imaginary part of the WDW equation gives the
continuity equation of the probability,
\begin{equation}
F\frac{\partial^2 S}{\partial a^2}
+ 2\frac{\partial F}{\partial a}\frac{\partial S}{\partial a} = 0 \, . \label{imag part}
\end{equation}

One may define the cosmological time via parameterizing the de
Broglie-Bohm trajectory along the tangential direction of the
gravitational action \cite{Banks:1985,Brout:1987ya,Brout:1987tq}:
\begin{equation}
\frac{\partial}{\partial t}
= - \frac{4\pi G}{3a}\frac{\partial S}{\partial a}
\frac{\partial}{\partial a} \, . \label{cos time}
\end{equation}
With this definition of time, the semiclassical Einstein
equation gives the modified Friedmann equation,
\begin{equation}
\left(\frac{\dot{a}}{a}\right)^2 + \frac{k}{a^2}
= \frac{\Lambda}{3} - \frac{32\pi^2G}{9a^4}V_q \, , \label{mod Fr}
\end{equation}
where the quantum potential for quantum corrections can be
given by inverting Eq.\,(\ref{imag part})
\cite{Kim:1996ae,Kim:1997te}
\begin{equation}
V_q = -\frac{G}{2} \left[
\left(\frac{\left(a\dot{a}\right)^{\cdot}}{2a\dot{a}^2}\right)^2
- \frac{1}{2\dot{a}}\left(
\frac{\left(a\dot{a}\right)^{\cdot}}{a\dot{a}^2}\right)^{\cdot}
\right] . \label{quant pot}
\end{equation}

A few remarks are in order. Firstly, the origin of quantum
potential and thereby quantum corrections is quantum
fluctuations of the spacetime geometry, being the scale factor
$a(t)$ here. Thus the WDW equation in this case describes only
the homogeneously and isotropically fluctuating geometry. Note
that massless scalar fields under the symmetry of homogeneity
and isotropy are stiff matter with the equation of state $w =
1$, and contribute to the modified equation (\ref{mod Fr}) as
\begin{equation}
\left(\frac{\dot{a}}{a}\right)^2 + \frac{k}{a^2}
= \frac{\Lambda}{3} -  \frac{32\pi^2G}{9a^4}V_q
+ \frac{4\pi G}{3a^6} \sum_{i = 0}^n p_i^2 \, , \label{Fr scalar}
\end{equation}
where $p_i$ is the momentum of the $i$-th scalar field.
Secondly, due to the quantum potential in the modified
Friedmann equation, the classical dS metric is no longer a
solution. It is extremely difficult to exactly solve the
nonlinear differential equation (\ref{mod Fr}). 
Nevertheless, when the quantum corrections are much smaller
than the cosmological constant, the scale factor $a(t)$ may
still approximately follow the dS behavior. Under this
approximation we have
\begin{equation}
V_q \simeq -\frac{G}{a^2} \, ,
\end{equation}
and the modified Friedmann equation reduces to
\begin{equation} \label{eq:WDW-Friedmann-eq-approx1}
\left(\frac{\dot{a}}{a}\right)^2 + \frac{k}{a^2}
\simeq H_{\Lambda}^2 + \frac{c_6}{a^6} \, ,
\quad c_6 \equiv \frac{32\pi^2}{9}G^2 \, .
\end{equation}
It is interesting to note that the quantum correction under the
dS approximation is proportional to $a^{-6}$ and therefore
behaves like a stiff matter as the homogeneous and isotropic
massless scalar fields in Eq.\,(\ref{Fr scalar}).

With the Friedmann equation (\ref{eq:WDW-Friedmann-eq-approx1})
we obtain the dS temperature via the previously demonstrated
method invoking the Euclidean time period. Thus the inverse temperature is given by
\begin{equation}
\beta_1 \equiv \frac{1}{k_B T_1} = \frac{2}{H_{\Lambda}}
\left( \int_{-\tilde{a}_1}^{-\tilde{a}_0} + \int_{\tilde{a}_0}^{\tilde{a}_1} \right)
\frac{d\tilde{a}}{\sqrt{1-\tilde{a}^2-(\tilde{c}_6/\tilde{a})^{4}}} \, ,
\end{equation}
where
\begin{equation}
\tilde{a} \equiv H_{\Lambda}a \, , \quad
\tilde{c}_6 \equiv H_{\Lambda} c_6^{1/4} \sim \frac{H_{\Lambda}}{M_\textrm{pl}}
\quad \textrm{($M_\textrm{pl}$: Planck scale),}
\end{equation}
are the dimensionless quantities and $\tilde{a}_{0}$ and $\tilde{a}_{1}$ are the only two positive roots of the
denominator of the integrand. When $\tilde{c}_6 \ll 1$,
\begin{equation} \label{eq:dS-Temperature1}
T_1 \simeq \frac{H_{\Lambda}}{2\pi k_B }
\left\{ 1 + \frac{\tilde{c}_6}{4\sqrt{2}\,3^{3/4}\,\Gamma\left(5/4\right)^2}
+ \mathcal{O}\left[\tilde{c}_6^2\right]
\right\}.
\end{equation}
We note that the leading correction proportional to
$\tilde{c}_6$ is positive and therefore $T_1 > T_0$.
Consequently the quantum correction makes the dS temperature
higher than that of classical spacetime in Eq.\
(\ref{eq:dS-Temperature0}).


\textit{de Sitter radiation in quantum cosmology with
inhomogeneous metric perturbations}.---
The dS radiation from the inhomogeneous perturbations
contribute to the Friedmann equation a relativistic energy
density ($\propto a^{-4}$) as follows,
\begin{equation} \label{eq:WDW-Friedmann-eq-approx2}
\left(\frac{\dot{a}}{a}\right)^2 + \frac{k}{a^2}
\simeq H_{\Lambda}^2
+ \frac{c_4}{a^4} + \frac{c_6}{a^6} \, ,
\end{equation}
where $c_4$ is an undetermined constant related to the
abundance of the dS radiation from inhomogeneous perturbations.
This can be read from Ref.\ \cite{Halliwell} where Halliwell
and Hawking investigated the inhomogeneous or isotropic modes
of a closed FLRW universe in the framework of quantum
cosmology. In the WDW equation, in addition to the
second-derivative terms deduced from the conjugate
momentum-square of these modes, the spatial variation of these
modes contributes the ``mass'' terms that are proportional to
$a^4$ and to the field-strength-square (see Eqs.\
(5.30)--(5.32) in Ref.\ \cite{Halliwell}). For example,
regarding the mode of the standard scalar harmonics on a
three-sphere with $a_{nlm}$ as the amplitude or the field
strength of the mode, in the WDW equation this mode contributes
a second-derivative term $\partial^2/\partial a_{nlm} {}^2$
from its conjugate momentum and a ``mass'' term $-a^4 (n^2 -
5/2) a_{nlm}^2 /3$ from its spatial variation. Such mass terms
will eventually contribute to the Friedmann equation an energy
term proportional to $a^{-4}$ like the $c_4$ term in Eq.\
(\ref{eq:WDW-Friedmann-eq-approx2}).

Via the Euclidean time period method, the Friedmann equation
(\ref{eq:WDW-Friedmann-eq-approx2}) gives the inverse dS temperature:
\begin{equation}
\beta_2 \equiv \frac{1}{k_B T_2} = \frac{2}{H_{\Lambda}}
\left( \int_{-\tilde{a}^*_1}^{-\tilde{a}^*_0} + \int_{\tilde{a}^*_0}^{\tilde{a}^*_1} \right)
\frac{d\tilde{a}}{\sqrt{1-\tilde{a}^2-(\tilde{c}_4/\tilde{a})^{2}-(\tilde{c}_6/\tilde{a})^{4}}} \, ,
\end{equation}
where $\tilde{a}^*_{0,1}$ are the only two positive roots of
the denominator of the integrand, and the dimensionless constant
$\tilde{c}_4 \equiv H_{\Lambda}c_4^{1/2}$.
Consider two limiting cases: $c_6$ domination and $c_4$
domination, regarding the quantum correction to the Friedmann
equation. Since the $c_6$ correction decreases faster than the
$c_4$ correction along with the cosmic expansion, at earlier
times the $c_6$ correction may dominate over $c_4$ and later
$c_4$ dominates. When $c_6$ dominates, i.e., when the
contribution from the homogeneous perturbation to the dS
radiation dominates,
we find $T_2 \simeq T_1$.
In contrast, later, when the contribution from inhomogeneous
perturbations dominates ($c_4$ domination), the dS temperature
approaches that of a classical spacetime: $T_2 \simeq T_0$.
We particularly note that, when $c_6=0$, we have $T_2 = T_0$,
regardless of the value of $c_4$, i.e., regardless the
abundance of the radiation. Thus, regarding the backreaction of
the dS radiation and the resultant quantum correction to the dS
temperature, the $c_4$-type radiation alone does not change the
dS temperature.
Between these two limiting cases, $T_0 < T_2 < T_1$. As a
result, along with the cosmic expansion, the dS temperature
decreases from $T_2$ to $T_0$.


Here we compare our work with the studies of the backreaction
of the dS radiation in the framework of a classical spacetime.
For example, Ref.\ \cite{Greene} investigates the backreaction
of the spherically symmetric dS radiation to dS space in the
Painlev\'{e} coordinates and finds that the dS temperature
decreases after the emission of the dS radiation, with the
decrement proportional to the energy of the radiation and to
$(H_{\Lambda}/M_\textrm{pl})^2$. For comparison, we note that
in our work the dS radiation is intrinsic quantum fluctuations
of gravity that exist at all times but no process of dS
radiation emission is involved. Along with the cosmic
expansion, the temperature of such purely gravitational dS
radiation decreases to $H_{\Lambda}/(2\pi)$ from a higher value
with the temperature difference that is caused by quantum
corrections and characterized by $H_{\Lambda}/M_\textrm{pl}$.

\textit{Summary}.---
We investigate the dS radiation and its backreaction to dS
space in quantum cosmology where the dS radiation is expected
to be inherent in the wave function of the universe. We
consider the simple system with pure gravity and a positive
cosmological constant. We have shown that the dS radiation from
the homogeneous perturbation contributes to the modified
Friedmann equation an energy density behaving like stiff
matter, while the dS radiation from inhomogeneous perturbations
may contribute a relativistic energy density. We have shown
that the backreaction of the dS radiation increases the dS
temperature in the case with a significant contribution from
the homogeneous perturbations. In contrast, the dS radiation
from inhomogeneous perturbations barely changes the dS
temperature, an interesting feature we particularly note.
The increment of the dS temperature is characterized by
$H_{\Lambda}/M_\textrm{pl}$, the ratio of the dS scale to the
Planck scale. Thus, when the dS scale approaches the Planck
scale, the temperature increment induced by the quantum
correction is significant. On the other hand, along with the
cosmic expansion the quantum correction is more and more insignificant
and the dS temperature decreases and approaches to $H_{\Lambda}/(2\pi)$,
the temperature of a classical dS space.


\textit{Acknowledgments}.---
J.~A.~G.\ would like to thank the hospitality of Kunsan
National University, where part of this paper was done.
S.~P.~K.\ would like to thank W-Y.~Pauchy~Hwang for the warmest
hospitality at Institute of Astrophysics, National Taiwan
University (NTU), where part of this paper was done. The work
of J.~A.~G.\ was supported by the National Science Council
(NSC) of Taiwan under Project No.\ NSC 98-2112-M-002-007-MY3.
The work of S.~P.~K.\ was supported in part by National Science
Council Grant (NSC 100-2811-M-002-012) of Taiwan and in part by
Basic Science Research Program through the National Research
Foundation of Korea (NRF) funded by the Ministry of Education,
Science and Technology (2011-0002-520). C.~M.~S.\ is supported
by Leung Center for Cosmology and Particle Astrophysics, NTU.

\end{document}